\documentclass{aa}
\begin{document}
\title{The Distance of the SNR Kes 75 and PWN PSR J1846-0258 System} 

\author{D.A. Leahy \inst{1}
\and 
     W.W. Tian  \inst{1,2}}
\authorrunning{Leahy \& Tian}
\titlerunning{The Distance of Kes 75/PWN PSR J1846-0258}
\offprints{tww@iras.ucalgary.ca. DAL and WWT give equal contributions to paper}
\institute{Department of Physics \& Astronomy, University of Calgary, Calgary, Alberta T2N 1N4, Canada;\\
\and
National Astronomical Observatories, CAS, Beijing 100012, China} 
\date{Received 27 Nov. 2007/Accepted 24 Jan. 2007} 

\abstract{The supernova remnant (SNR) Kes 75/PSR J1846-0258 association can be regarded as a certainty due to the accurate
 location of young PSR J1846-0258 at the center of Kes 75 and the detected bright radio/X-ray 
synchrotron nebula surrounding the pulsar. 
We provide a new distance estimate to the SNR/pulsar system by analyzing the HI and $^{13}$CO maps, 
the HI emission and absorption spectra, and the $^{13}$CO emission spectrum of Kes 75. 
That there are no absorption features at negative velocities strongly argues against the widely-used 
large distance of 19 to 21 kpc for Kes 75, and shows that Kes 75 is within the Solar circle, 
i.e. a distance $d<$13.2 kpc. Kes 75 is likely at a distance of 5.1 to 7.5 kpc because the
 highest HI absorption velocity is at 95 km/s, and no absorption is associated with a 
nearby HI emission peak at  102 km/s in the direction of Kes 75. 
This distance to Kes 75 gives a reasonable luminosity of PSR J1846-0258 and its PWN and also leads to a much 
smaller radius for Kes 75, so the age of the SNR is consistent with the 
spin-down age of PSR J1846-0258, confirming this pulsar as the second-youngest in
the Galaxy. 

}
\maketitle

\section{Introduction}
The supernova remnant (SNR) Kes 75 (G29.7-0.3) is one of a few examples of a shell-type 
remnant harboring an extremely young pulsar that powers an extended radio/X-ray synchrotron
nebula. Ever since the young energetic X-ray pulsar, PSR J1846-0258, was discovered by Gotthelf 
et al. (2000) using the Rossi X-Ray Timing Explorer, many observations of this system at different wave-bands have been executed (Sugizaki et al. 2001, Mereghetti et al. 2002, Bock \& Gaensler 2005, Djannati-Ata\"i 2007, McBride et al. 2008). 
 These observations strongly support the SNR/pulsar association  
because of both the position of PSR J1846-0258 at the geometrical centers of the SNR shell and 
the bright radio/X-ray nebula within the SNR. 

A young pulsar usually favors a small radius of the associated SNR, but Kes 75 had an 
estimated radius of 10 pc based on the previously estimated 19 kpc distance. 
Caswell et al. (1975) employed the Parkes HI-line interferometer to find the most distant 
reliable absorption feature at velocity of $\sim$ 90 km/s in the direction of Kes 75, and 
suggested the lower-limit distance of 6.6 kpc (assuming Solar - Galactic center distance R$_{0}$= 10 kpc). This drops to
5.0 kpc using a recent measurement of R$_{0}$= 7.6$\pm$0.3 kpc (Eisenhauer et al 2005). Subsequent distance estimates
of 19 to 21 kpc for Kes 75 were obtained. 
Becker \& Helfand (1984) suggested $\sim$ 21 kpc using VLA (D array) HI absorption spectrum observations with features at negative velocity.
Based on d=19 to 21 kpc, the extremely high X-ray luminosities for Kes 75 
($\sim$ 4$\times10^{37}$ ergs s$^{-1}$), the pulsar, and the PWN require a highly implausible 
explosion energy ($\sim$ 10$^{53}$ erg), leading to several progenitor evolutionary scenarios 
(Morton et al. 2007, Helfand et al. 2003). The most likely one of the scenarios points to a 
large error in the distance estimate of $\sim$ 19 kpc (Morton et al. 2007). 

The 21 cm neutral hydrogen absorption measurements to the Galactic objects can generally provide a 
distance estimate with an uncertainty of $<$1 kpc. Recently we used  HI+$^{13}$CO 
observations to measure and revise distances to a few SNRs with high precision 
(Leahy \& Tian 2008, Tian et al. 2007, Tian \& Leahy 2008).  In this paper, we significantly revise the 
distance to Kes 75 using these methods. 
The radio data come from 1420 MHz continuum plus 
HI-line observations of the VLA Galactic Plane Survey (VGPS) and  the $^{13}$CO-line (J = 1-0) 
observations of the Galactic ring survey 
(Stil et al. 2006, Jackson et al. 2006).

\begin{figure*} 
\vspace{50mm} 
\begin{picture}(50,50) 
\put(-80,230){\includegraphics{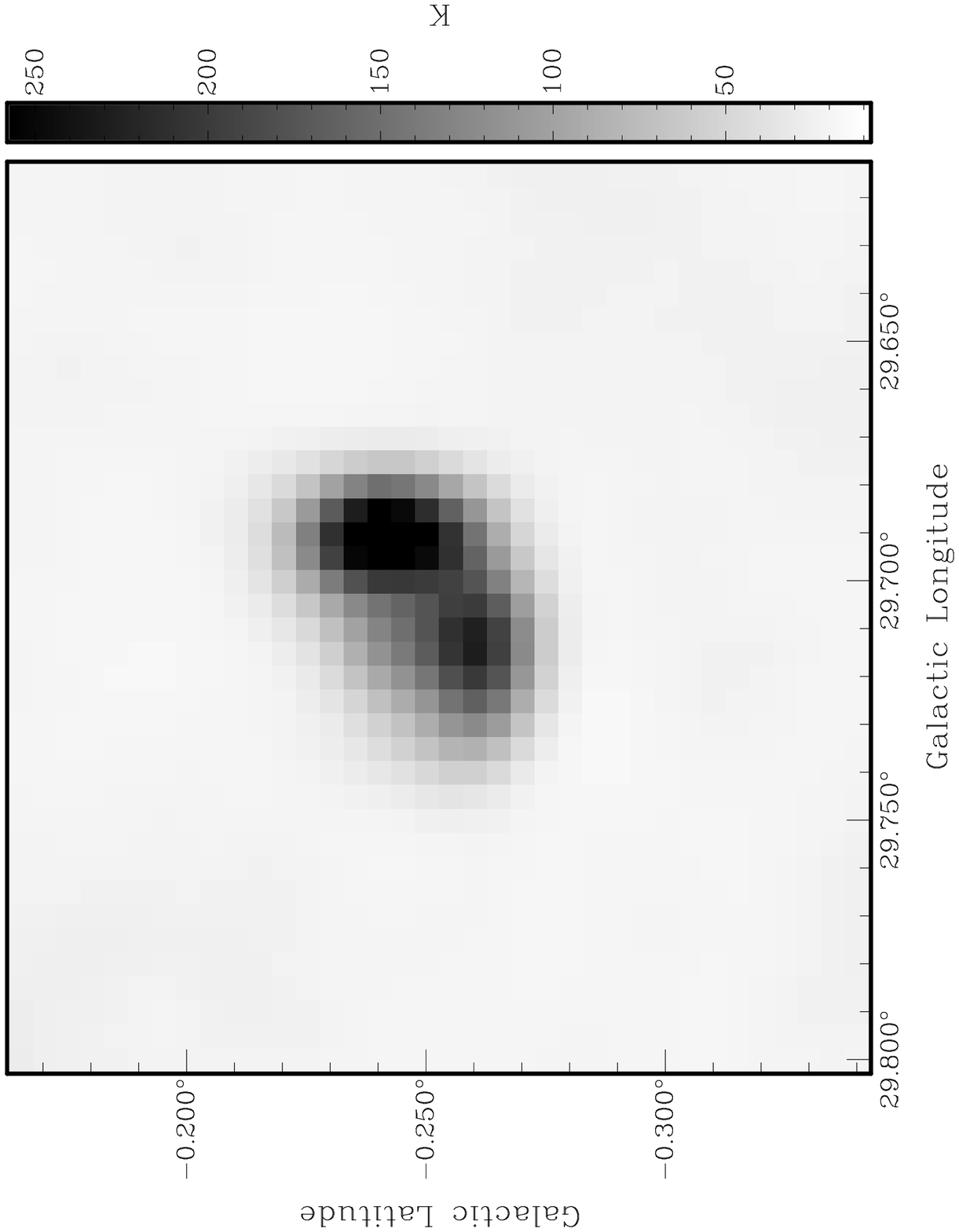}} 
\put(240,-20){\includegraphics{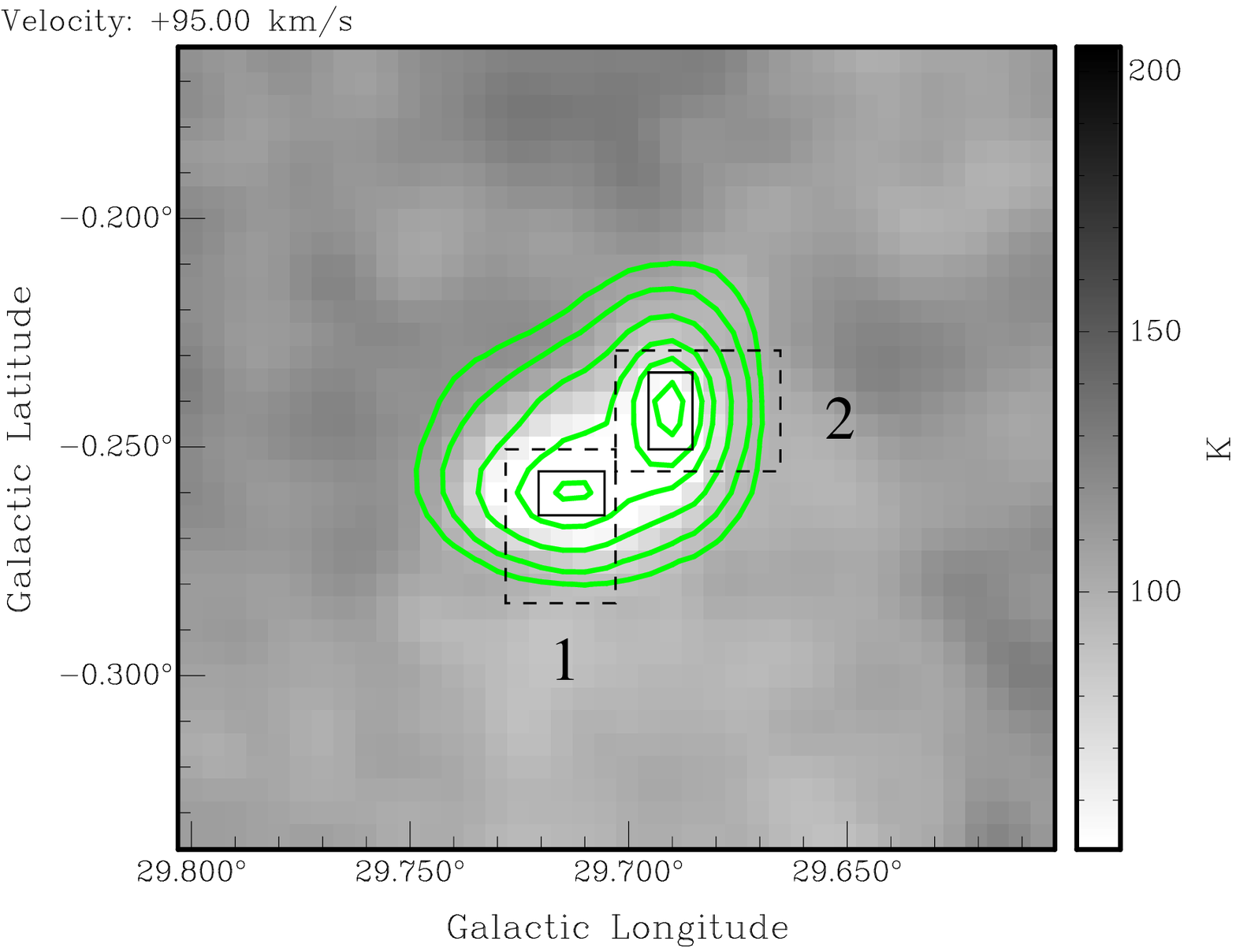}} 
\end{picture} 
\caption{The 1420 MHz continuum image (left) and HI image (right) of Kes 75 from a single channel. The HI map has superimposed contours (28, 40, 60, 100 K) of the 1420 MHz continuum emission to show the SNR.} 
\end{figure*} 

\begin{figure*}
\vspace{160mm}
\begin{picture}(160,160)
\put(-55,645){\includegraphics{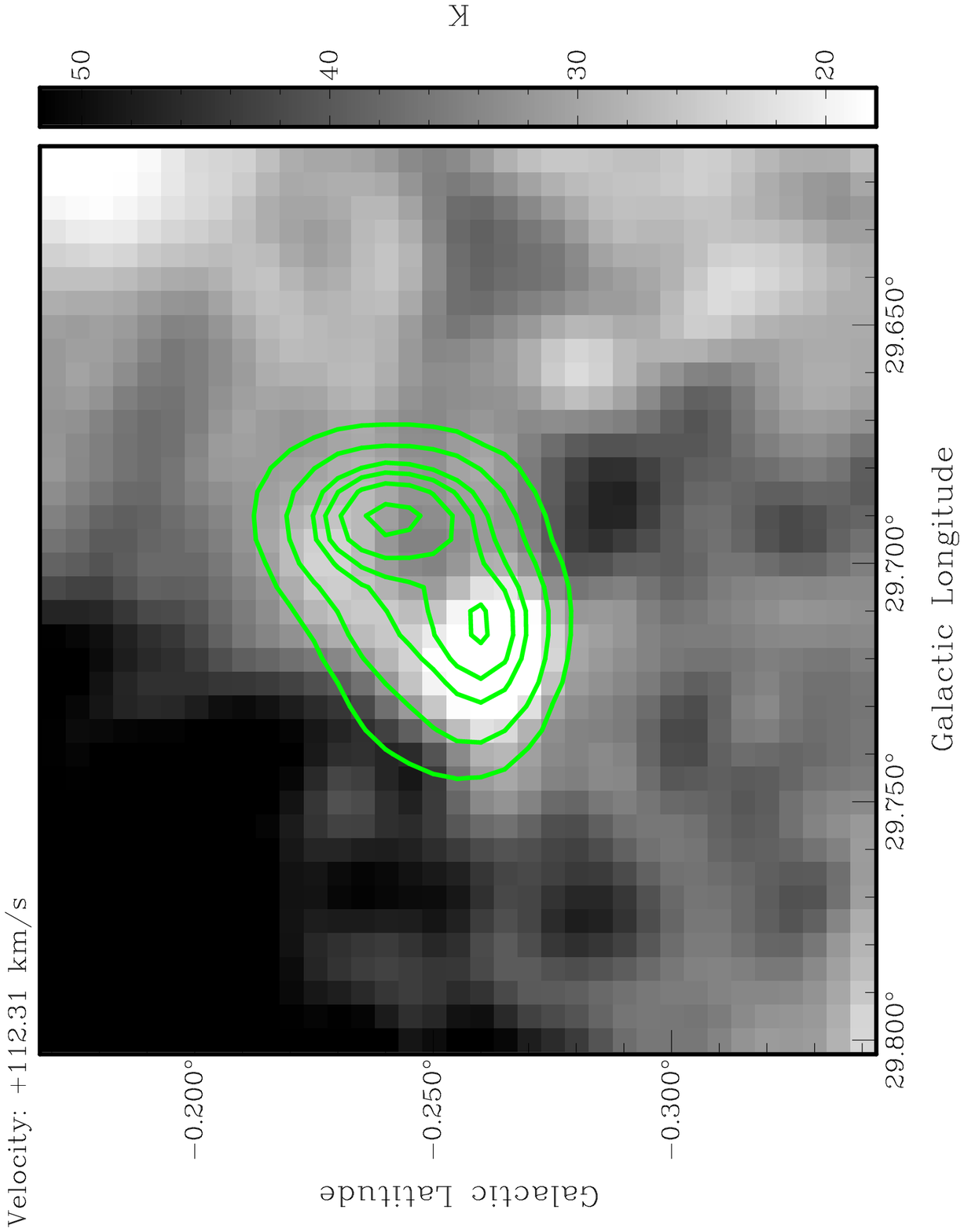}}
\put(230,375){\includegraphics{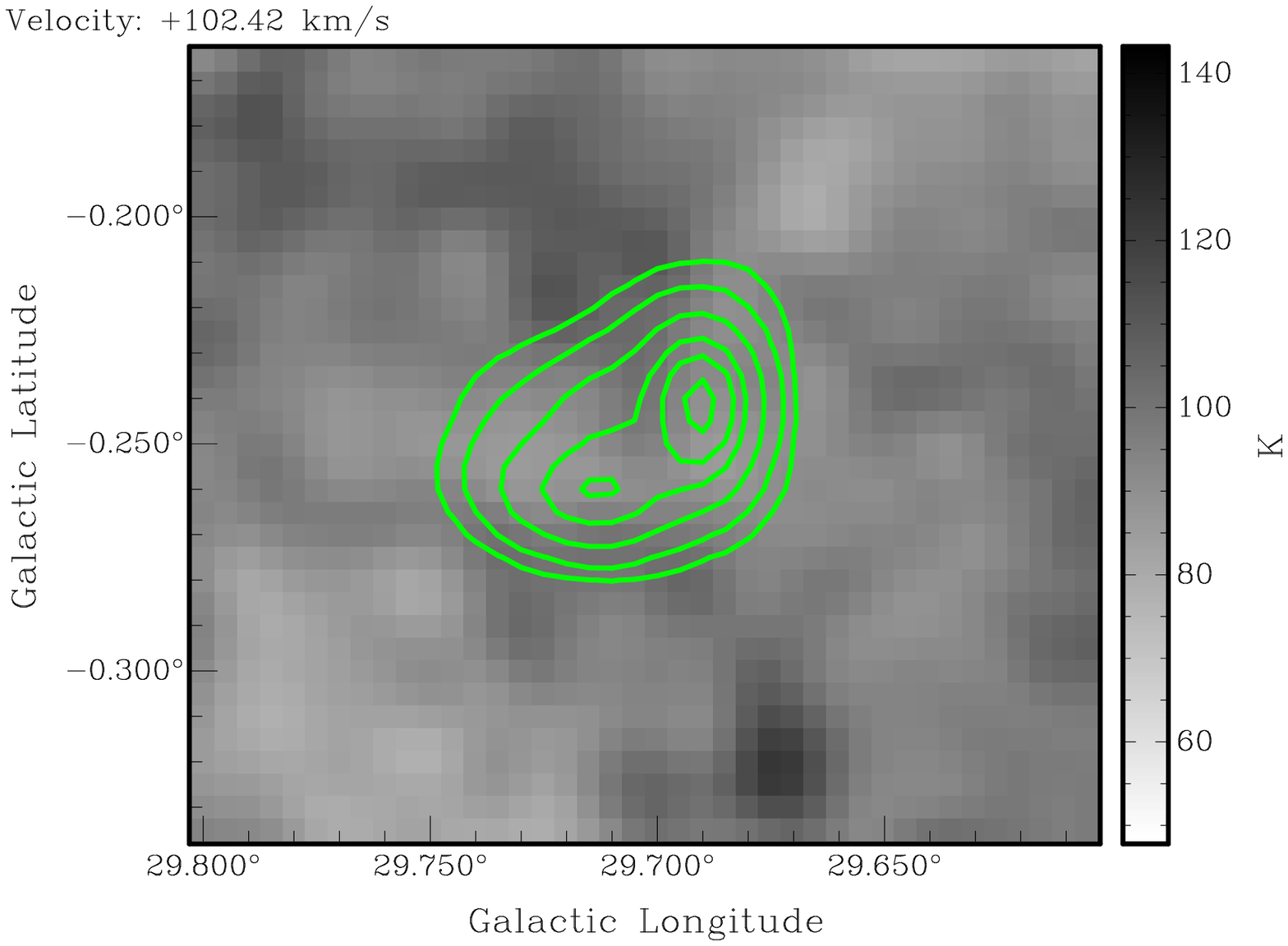}}
\put(-20,308){\includegraphics{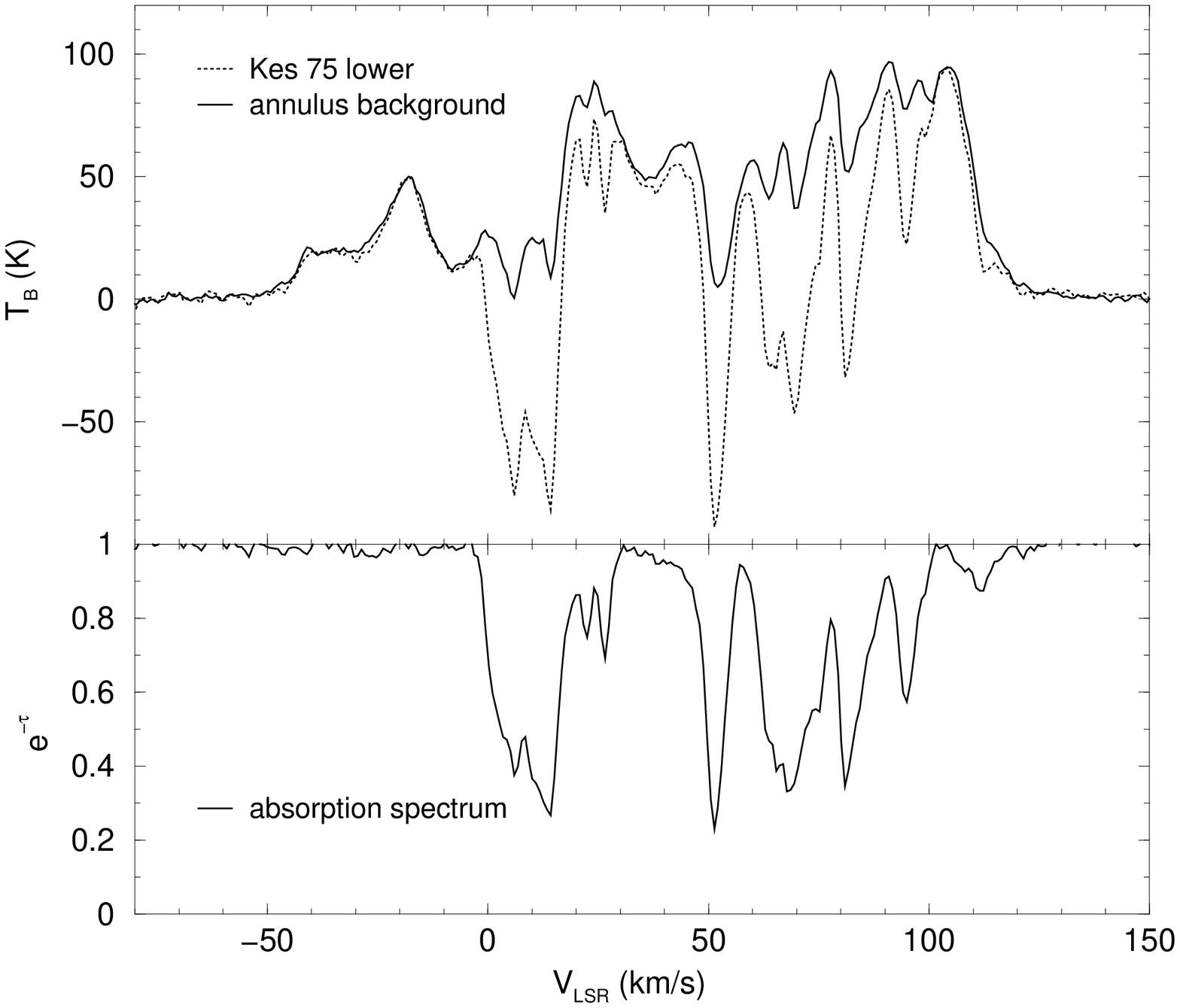}}
\put(230,478){\includegraphics{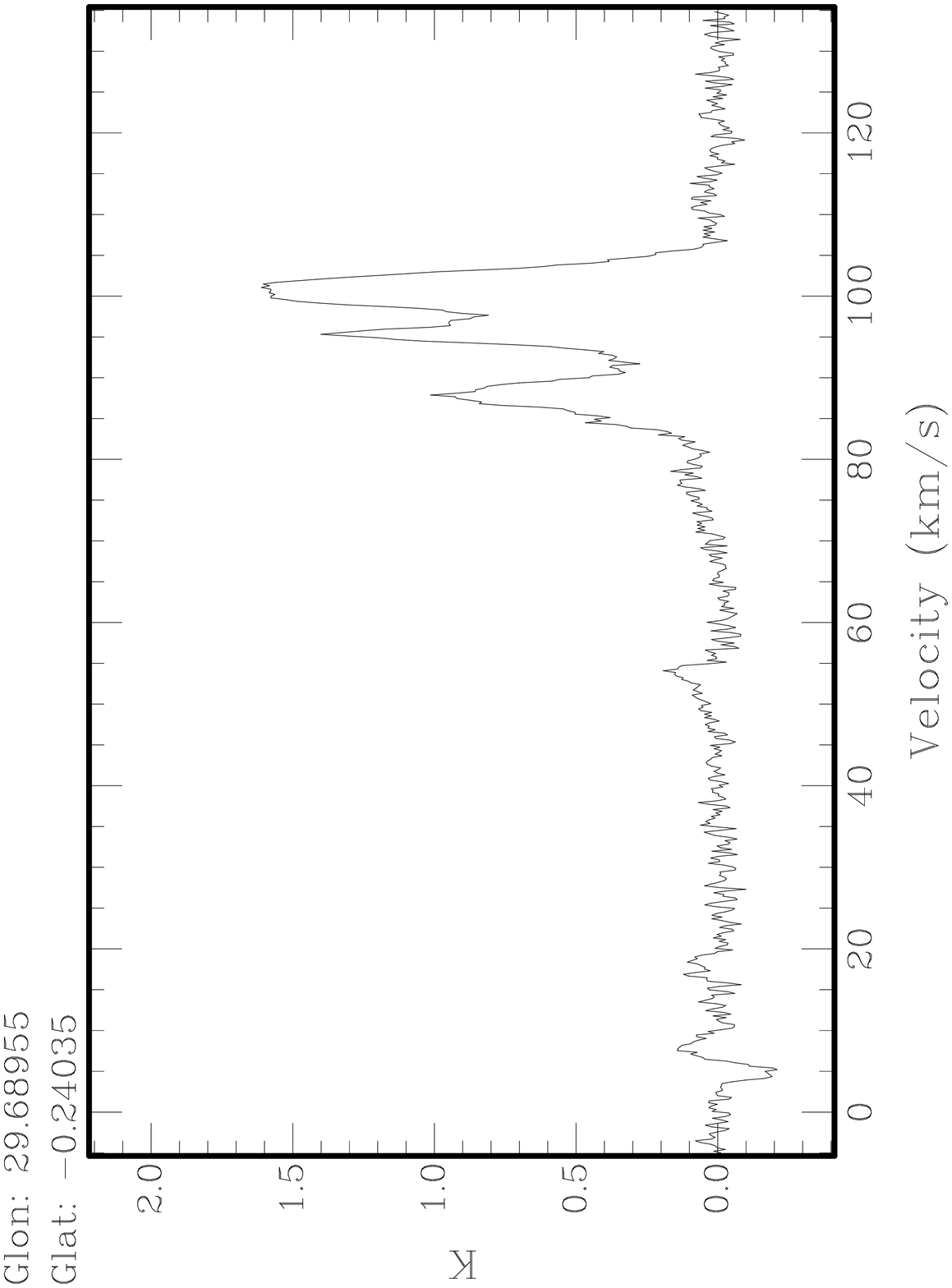}}
\put(-20,155){\includegraphics{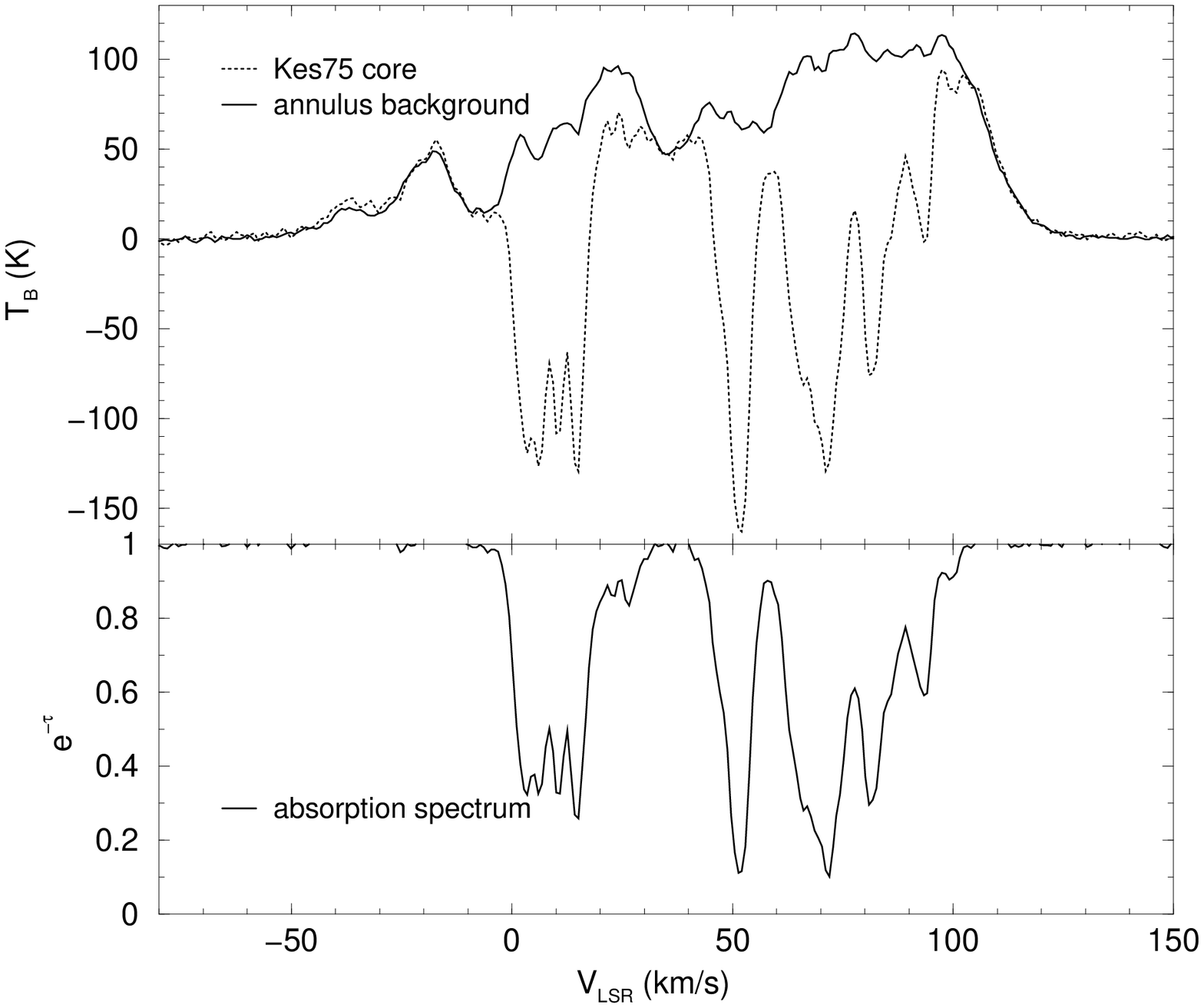}}
\put(230,325){\includegraphics{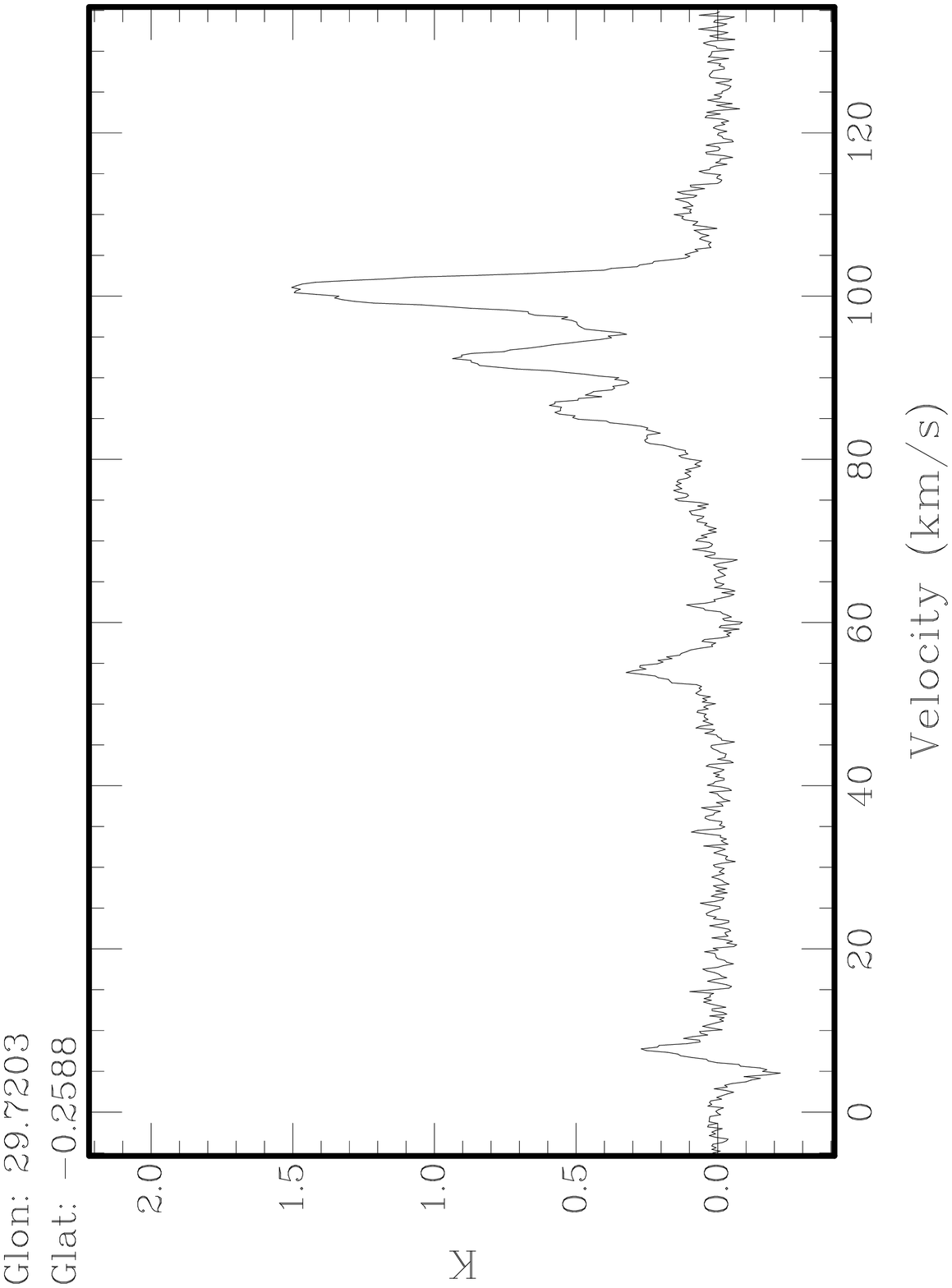}}
\put(-20,-10){\includegraphics{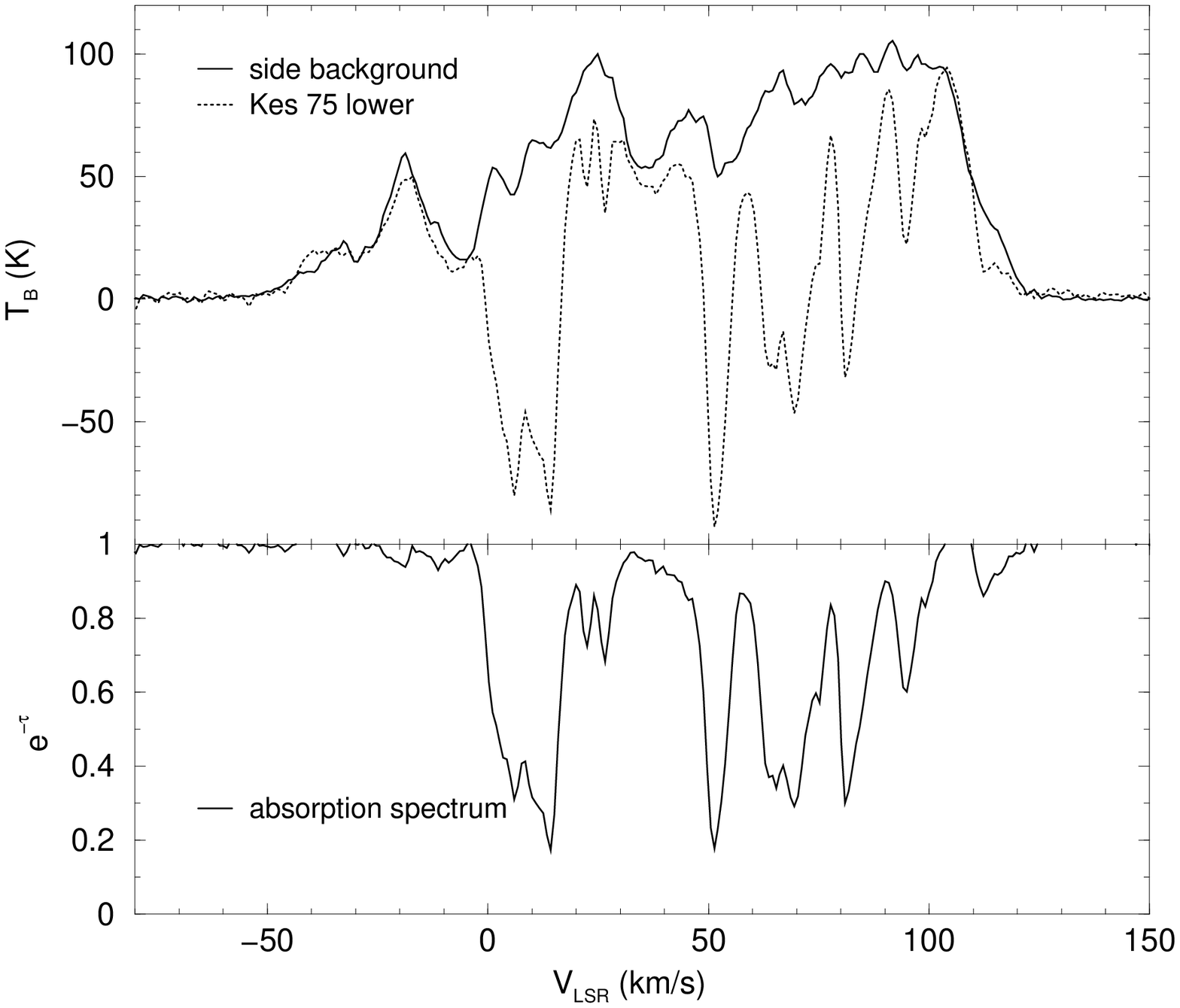}}
\put(200,180){\includegraphics{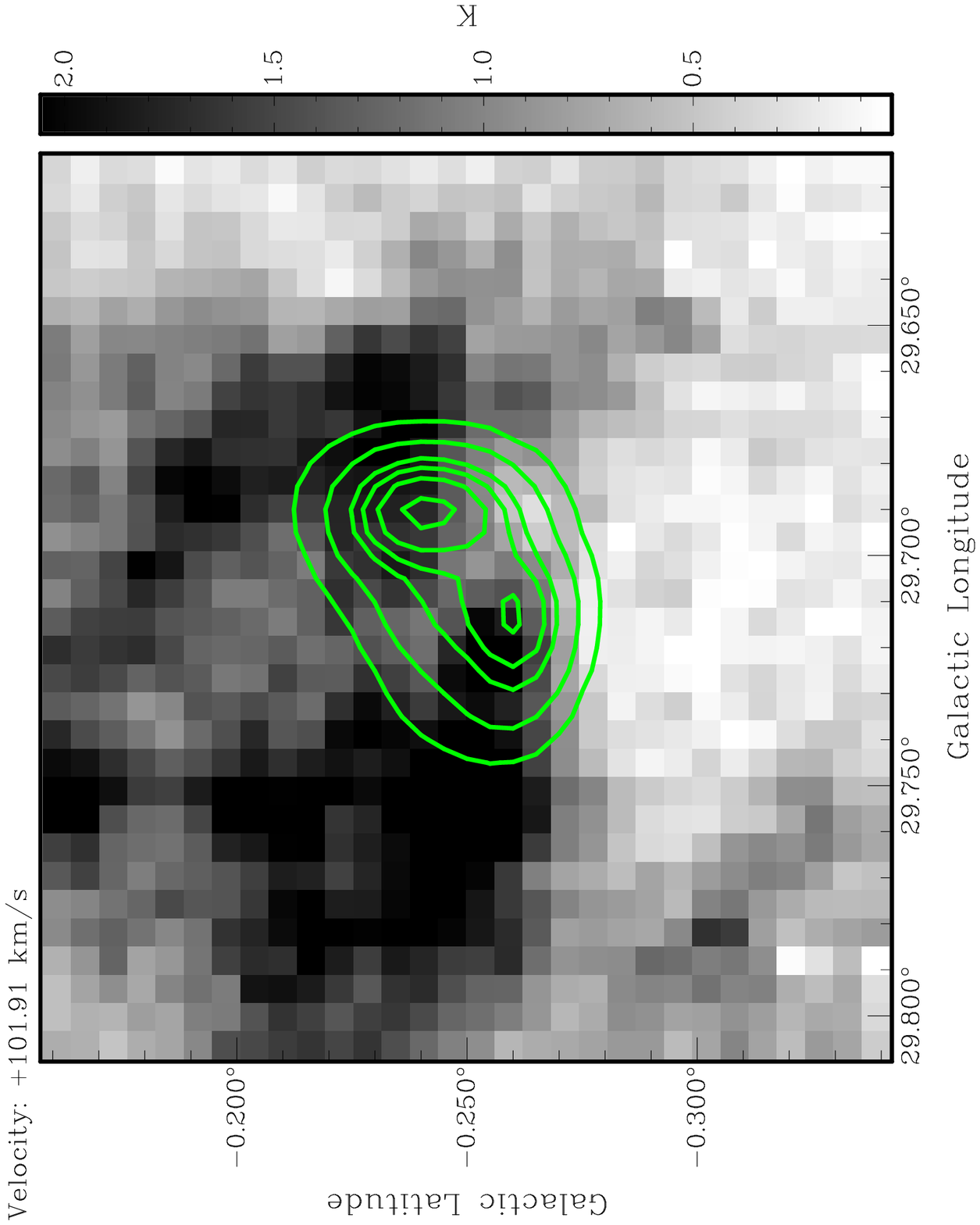}}
\end{picture}
\caption{Two channel HI images of Kes 75 (top row). The second and third rows show the HI spectra (left) and the CO emission spectra (right), extracted from boxes 1 and 2 shown in Fig. 1. The bottom row shows an additional HI spectra(left) with the source spectrum from box 1 and background spectrum from the area at $\l$[29.705, 29.725],$b$[-0.285, -0.310], and one-channel CO image (right). Both the HI and CO maps have superimposed the same contours as Fig. 1.}
\end{figure*}

\section{Analysis and results}

In Fig. 1, the VGPS 1420 MHz continuum and one-channel HI images of Kes 75 are given in Galactic coordinates, 
showing the two emission peaks from the SNR shell. The shell is shown at higher resolution 
using VLA in B, C, and D array by Helfand et al. (2006). 
Figure 2 shows two HI-channel maps, one CO-channel map, and the HI absorption/emission and CO emission spectra. 
The HI spectra do not show any absorption at negative velocities, so Kes 75 must lie within 
the Solar circle, which is at d$\simeq$ 13.2 kpc at $l=29.7^\circ$. 
This result 
significantly revises the previously widely-used value of 19 to 21 kpc for Kes 75. 
Here we use a circular rotation model with R$_{0}$=7.6 kpc and constant circular velocity V$_{0}$=220 km/s (1985 IAU standard value), yielding a tangent point velocity at $l=29.7^\circ$ of 111 km/s. This is consistent with the observation value in the direction of Kes 75 of $\sim$ 110 km/s from Fig. 2. 
We have also examined the HI channel images in the negative velocity range where Becker \& Helfand (1984) find the absorption features, and found no evidence of absorption. 

Further information on the distance of Kes 75 can be obtained by examining the absorption
spectra and HI and CO channel images. At 95 km/s there is clear absorption seen in the
HI channel map (Fig. 1, right panel) and in the absorption spectra of both continuum peaks (boxes 1 and 2 in Fig. 2) of Kes 75. 
To verify our method of using annulus background, we calculated an HI absorption spectrum of box 1 using an alternate background (see Fig. 2 lower left panel) and obtain almost identical spectra. Ninety-five km/s is the highest velocity of prominent HI absorption appearing in the spectra (or channel images) of Kes 75, and 
it corresponds to a minimum distance of 5.1 kpc. 
Also, there is no absorption in the HI channel map (Fig. 2 top right panel)
or in the HI absorption spectra 
for the HI emission peak at 102 km/s in the direction of Kes 75.
Thus Kes 75 is nearer than this emission peak,
which could be at either the near side or the far side of the tangent point. 
If this gas is on the near side, the upper limit for the distance of Kes 75 is 5.5 kpc;
in the case that HI is on the far side, the upper limit is 7.5 kpc. 
We used the latter in the absence of evidence that this emission peak is on the near side of the tangent point.
The CO emission spectra in the direction of boxes 1 and 2 both show a high 
brightness-temperature cloud component at a velocity of 102 km/s. 
Since there is no strong respective HI absorption in the spectra of Kes 75, the CO cloud 
should be behind Kes 75 and is probably
associated with the HI-emission peak at 102 km/s.
Thus both HI and CO observations show that Kes 75 is nearer than the interstellar
gas at 102 km/s and confirm an upper distance limit of 7.5 kpc. 

There is one puzzling feature that seems to be caused by true absorption at $\sim$112 
km/s in the direction of box 1, but not in the direction of box 2 (Fig. 2, HI channel map, HI and CO spectra). 
This cannot be due to diffuse HI, or else it would appear in absorption against both continuum
peaks of Kes 75. This feature is probably caused by a compact HI cloud at 112 km/s velocity.
Since the cloud is compact, it could either be at the tangent point or it could be moving at a peculiar velocity, and elsewhere along the line-of-sight. 
If the cloud has no peculiar velocity it provides a lower limit to the
distance to Kes 75 of the tangent point distance (6.6 kpc) and also confirms that the
HI at 102 km/s is on the far side of the tangent point. If it does have a peculiar
velocity, then this cloud gives no useful distance information.

In summary, Kes 75 should be in the distance range of 5.1 to 7.5 kpc  because of the 
absorption at 95 km/s and absence of absorption at 102 km/s.

\section{Discussion}  

We find strong evidence about the distance to Kes 75, and if disagrees with what is given
by Becker \& Helfand et al (1984). 
By testing different background regions, we 
were not able to obtain similar HI absorption spectra to those 
shown in that study. 
The VGPS data used here is superior to their data in
sensitivity, and it also includes single-dish observations from the 100 m Green Bank telescope, 
yielding the large-scale HI
and continuum emission in addition to small-scale emission. 
This is probably one reason we get a different, better, result.

Next we discuss implications of the shorter distance to Kes 75.
The new distance gives a much smaller radius for Kes 75: the radio and X-ray sizes nearly
match (Morton et al. 2007) and give a radius of 1.60 arcmin. Our new distance range
gives a physical radius range of 2.37 to 3.49 pc. We write the distance to
Kes 75 as $d=d_6 \times$6 kpc, so the radius is 2.80$d_6$ pc.
As estimate of the age of the SNR, we use the spin-down age of 723 years for PSR J1846-0258
from Gotthelf et al (2000). 
Livingstone et al (2006) use measurements of braking index of PSR J1846-0258 
to put an upper limit on age of 884 yr, confirming the spin-down age estimate. 
Assuming free expansion, the derived expansion velocity is then 3800$d_6$ km/s
with this spin-down age and a radius of 2.80$d_6$ pc. 
If the SNR is expanding according to Sedov evolution, the current Sedov shock velocity is
0.4 times the average expansion velocity, giving a shock velocity of 1500$d_6$ km/s. 
For the more realistic case, the initial phases
will be dominated by free-expansion, so the current shock velocity will be higher than
the Sedov estimate and lower than the free expansion estimate.
However, we note that Helfand et al. (2006) estimate the expansion velocity
of Kes 75 using the Si X-ray line width to obtain $\sim$3700 km/s.
This indicates that Kes 75 is probably closer to the free-expansion phase.

The electron temperature of the high-temperature component
in the X-ray spectrum is 1.4-1.5 keV (Morton et al. 2007), which corresponds to a
shock velocity (for electron-ion equilibration) of 1200 km/s. Since the minimum shock
velocity is 1500$d_6$ km/s, there must not be full electron-ion equilibration:
one finds a maximum electron-to-ion temperature ratio of 0.64.
The shock velocity derived from the electron temperature increases in this case. For
an electron-to-ion temperature ratio of 0.1, one obtains a shock velocity of 3800 km/s.
Thus the electron-to-on temperature ratio is restricted to the range
of 0.10 to 0.64 for Kes 75.

Morton et al (2007) give X-ray emitting gas masses in hot (1.4 keV)  and cool (0.23 keV) components. For our new distance, the masses are reduced to reasonable values: $5.3 f_1^{1/2}d_6^{5/2}M_{\odot}$ for the cool component and  
$0.84 f_1^{1/2}d_6^{5/2}M_{\odot}$ for the hot component with $f_1$ the filling factor. 
They interpret the cool component as shocked interstellar medium (ISM) or circumstellar 
 material (CSM) and the hot component as reverse-shocked ejecta. 
That two temperature components are seen is clear evidence that the ejecta has
not fully entered the Sedov phase.
The shocked-ejecta mass is likely to be smaller than the ISM/CSM shocked mass since only 
a small part of the ejecta has been shocked so far.
The expansion should have already slowed down somewhat from the initial ejection velocity, 
so the current shock velocity and electron-to-ion temperature ratio should be somewhere between the Sedov estimates (1500$d_6$ km/s and 0.64) and the free-expansion values (3800$d_6$ km/s and 0.10). 

Using the Sedov model and age of 723 yr, the observed outer shock radius yields 
an estimate of the explosion energy ($E_0$) divided by interstellar density ($n_0$): 
$E_0/n_0=0.69d_6^5\times10^{50}$erg.  The x-ray spectral fitting yields an estimate
of the density for the shocked CSM/ISM component of $n=100f_1^{-1/2}d_6^{-1/2}$cm$^{-3}$
and $n=64f_1^{-1/2}d_6^{-1/2}$cm$^{-3}$ for the SE and SW limbs of Kes 75, respectively.  
From the inferred X-ray post-shock density ($n$) and using a compression rate of 3 (from the observed radio spectral index of 0.7, see Droge et al. 1987), we obtain a preshock density $n_0=n/3\simeq 27 f_1^{-1/2}d_6^{-1/2}$ cm$^{-3}$ and
an explosion energy of $E_0=1.9d_6^{9/2}f_1^{-1/2}\times10^{51}$erg. This is entirely
consistent with what is expected for a core-collapse explosion now, contrary to the problems
caused by having a large distance to Kes 75. 

For PSR J1846-0258, the pulsar-plus-nebular luminosity in the 3-10 keV band is
$7.8 \times 10^{34} d_6^2$erg s$^{-1}$, so the pulsar-plus-nebula luminosity to spin-down
luminosity has a slightly lower ratio (0.6$d_6^2$) than for the Crab, rather than 
the previously inferred, much higher ratio.

In summary, we find observational evidence that the distance to Kes 75 is about 6 kpc and that it has a normal explosion energy and harbors a Crab-like rotation powered pulsar
and pulsar wind nebula. This is unusual in  that the pulsar has a high value for magnetic field ($5\times10^{13}$ Gauss),  and it is very young. It also exploded in a somewhat high-density environment: $n_0\simeq$27 cm$^{-3}$.

We can estimate the absorbing column density to Kes 75 $N_{HI}$ from the HI absorption spectrum (Fig. 2) by $N_{HI}$=1.9$\times$10$^{18}$$\tau$$\Delta{\it{v}}$$T_{s}$ cm$^{-2}$ (Dickey \& Lockman 1990). Assuming HI spin temperature $T_{s}$ of 50K, we sum over the absorption features to obtain $N_{HI}$ of $\sim$ $3\times$10$^{22}$ and $\sim$ $4\times$10$^{22}$ cm$^{-2}$ in front of regions 1 and 2, respectively. These values agree with that of $N_{H}$=4$\times$10$^{22}$ cm$^{-2}$ derived from X-ray observations to Kes 75.
Given the high column density to Kes 75, the 
derived extinction is $A_V \sim 21$mag, which would explain why no historical SN has been observed 
associated with Kes 75 around the time of explosion in 1300 A.D.

\begin{acknowledgements}
DAL and WWT acknowledge support from the Natural Sciences and Engineering Research Council of Canada. WWT acknowledges support from the Natural Science Foundation of China.  
This publication makes use of molecular line data from the Boston University-FCRAO Galactic Ring Survey. The NRAO is a facility of the National Science Foundation operated under cooperative agreement by Associated Universities, Inc. 
\end{acknowledgements}

\end{document}